\documentclass[aps,prl,showpacs,twocolumn,superscriptaddress]{revtex4}

\usepackage{amsmath,amssymb}
\usepackage{graphicx}
\usepackage{psfrag,color}
\usepackage{cleveref}

\begin{document}

\title{Transition from viscous to inertial regime in dense suspensions}
\author{Martin Trulsson} \email{martin.trulsson@espci.fr} \author{Bruno Andreotti} \author {Philippe Claudin} 
\affiliation{Physique et M\'ecanique des Milieux H\'et\'erog\`enes, UMR 7636 ESPCI -- CNRS -- Universit\'e Paris-Diderot -- Universit\'e P.M.~Curie, 10 rue Vauquelin, 75005 Paris, France}

\begin{abstract}
Non-Brownian suspensions present a transition from Newtonian behaviour in the zero-shear limit to a shear thickening behaviour at a large shear rate, none of which is clearly understood so far.  Here, we carry out numerical simulations of such an athermal dense suspension under shear, at an imposed confining pressure. This setup is conceptually identical to the recent experiments of Boyer and co-workers \cite{BGP11}. Varying the interstitial fluid viscosities, we recover the Newtonian and Bagnoldian regimes and show that they correspond to a dissipation dominated by  viscous and contact forces respectively. We show that the two rheological regimes can be unified as a function of a single dimensionless number, by adding the contributions to the dissipation at a given volume fraction.
\end{abstract}

 \date{\today}

\maketitle

The rheology of amorphous materials such as emulsions, foams, metallic glasses, suspensions or granular materials share a similar phenomenology close to the jamming transition at which viscosity diverges \cite{BGP11,Tighe10,Durian95,Olsson07}. However, the dynamics of these systems is not yet clearly understood and the establishment of a unified theory remains a challenging goal of out of equilibrium statistical physics. Following the pioneering work of Einstein  \cite{Einstein05}, the common view on suspensions of particles in a fluid has long been to start from the dilute limit and to perform an expansion in volume fraction $\phi$ \cite{Batchelor77,BookGuazzelli}, with a particular emphasis on the effective interaction between particles mediated by the fluid. By contrast, recent studies have started to view the rheology of dense suspensions from the other limit instead, in the framework of dense granular systems \cite{Lemaitre09,BGP11,Fall2010,MS09,Lerner12,Andreotti12}. The rheology of dense suspensions of solid particles in an isodense fluid of viscosity $\eta_f$ is Newtonian at small shear rate $\dot \gamma$ with a viscosity $\tau/\dot \gamma$ diverging as $\eta_f (\phi_c-\phi)^{- \beta}$, as the particle volume fraction goes to its critical value $\phi_c$. The measured exponent $ \beta$ ranges between $2$ and $3$ \cite{Zarraga01,Ovarlez06,BDCL10,BGP11}. Mean field theory assuming a dissipation dominated by lubrication films separating particles predicts an exponent $ \beta=1$ \cite{MS09}. By contrast, numerical simulations assuming that dissipation is due to the nonaffine displacement of particles give the exponent $ \beta\simeq 2.2$ \cite{Lerner12,Andreotti12}. There they relate the zero-shear viscosity, a macroscopic dynamical observable, to a microscopic observable: the variance of the nonaffine velocity/displacement \cite{Andreotti12}. The latter is itself related to the geometry of the contact network \cite{Lerner12}.

While most fluids shear thin, it was first shown by Bagnold~\cite{B54} that suspensions exhibit shear {\it thickening} when the volume fraction $\phi$ is kept constant: their apparent viscosity increases with the shear rate. However, the conditions  for such a property to emerge still remain controversial~\cite{Fall2010}. In particular, as recently emphasized \cite{BGP11}, suspensions exhibit shear {\it thinning} when the confining pressure $P^p$ is controlled and kept constant, a property reminiscent of dry granular materials. 
\begin{figure}[t!]
\begin{center}
\includegraphics[scale=1]{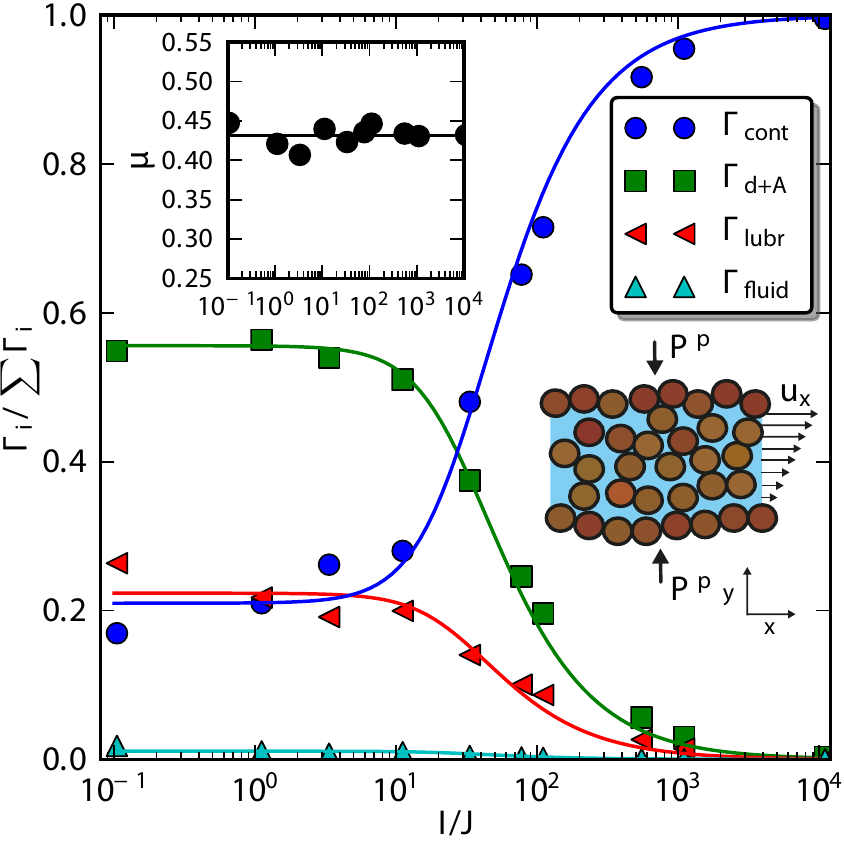}
\vspace{0 mm}
\caption{Fraction of the power dissipated by contact forces ($\Gamma_{\rm cont}$), viscous drag plus Archimedes forces ($\Gamma_{\rm d+A}$), lubrication forces ($\Gamma_{\rm lubr}$) and fluid viscosity ($\Gamma_{\rm fluid } =\eta_f \dot{\gamma}^2$), as a function of the ratio $I/J$ for a fixed value of the volume fraction ($\phi \simeq 0.78$).  Solid lines are the best fits to the expression $\frac{c_i J}{J+\alpha I^2}$, with $c_i$ as fitting parameters, for the three last dissipation components. Insets: Friction coefficient $\mu$ against $I/J$ (solid curve illustrates the average $\mu$) and a schematic of the numerical setup.}
\label{fig:DispM}
\vspace{0 mm}
\end{center}
\end{figure}

In this Letter, we use discrete element simulations of non-Brownian particles interacting with a continuum viscous fluid to show that the rheology of suspensions at finite shear rate can be unified with the Newtonian quasistatic limit. More precisely, Boyer \textit{et al.} \cite{BGP11} have recently shown that the rheology of a suspension approaching the zero-shear limit can be rewritten as a frictional law of the form $\tau=\mu_J(J) P^p$ and $\phi=\phi_J(J)$, where $J=\eta_f \dot \gamma/P^p$ is the viscous number  comparing viscous stresses to the confining pressure. In the inertial Bagnoldian regime, the flow is characterized by  the inertial number $I=\sqrt{\rho \dot \gamma^2 d^2/P^p}$, with a subsequent rheology of the form $\tau=\mu_I(I) P^p$ and $\phi=\phi_I(I)$. We show here that the contributions to the dissipation can be added at fixed $\phi$, which results in a unique rheology $\tau=\mu(K)$ and $\phi(K)$ controlled by the dimensionless number $K=J+\alpha I^2$, where $\alpha$ is a constant of order $1$ encoding the details of dissipative mechanisms.
\begin{figure}[t!]
\begin{center}
\includegraphics[scale=1]{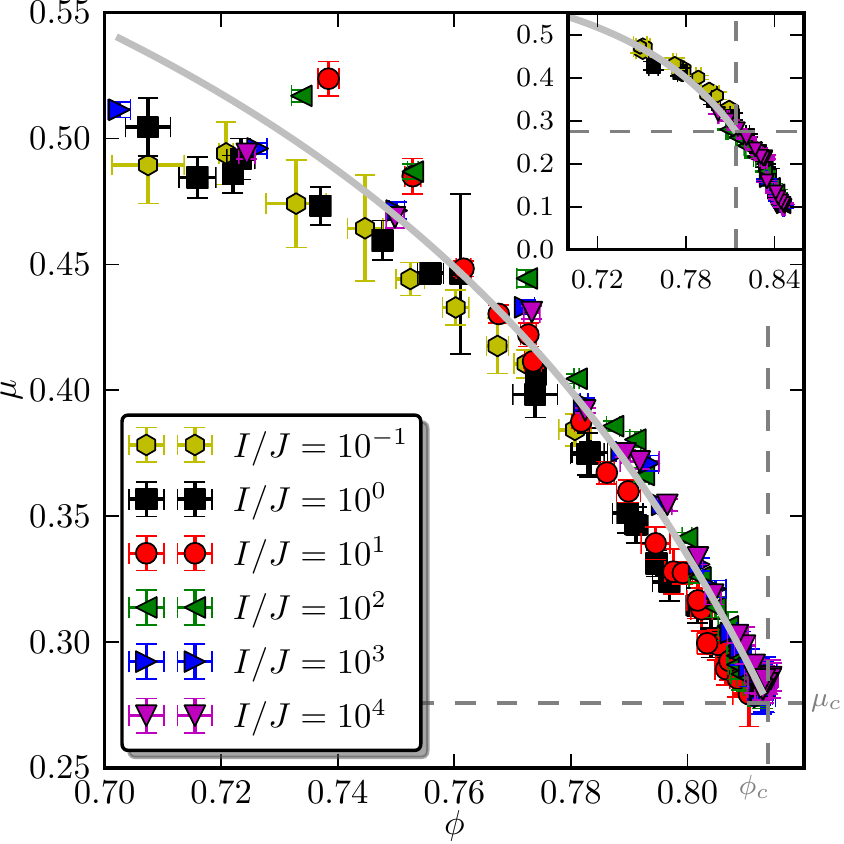}
\vspace{0 mm}
\caption{Friction coefficient of the suspension $\mu=\tau/P^p$ as a function of the particle volume fraction $\phi$, for different values of $I/J$. Inset: Same as the main plot but for frictionless grains ($\mu_p=0$). The solid lines are the best fit by Eq.~(\ref{eq:muphi}) for $\mu_p=0.4$ and the dashed show the critical values for
$\mu_p=0.4$.}
\label{fig:mu_phi}
\vspace{0 mm}
\end{center}
\end{figure}
 
{\it Numerical model.~--} We consider a two-dimensional system constituted of  $\simeq 10^3$ circular particles of mass $m_i$ and diameter $d_i$, with a $\pm$50 \% polydispersity. The shear cell is composed by two rough walls, created by gluing together two dense layers of grains, with periodic boundary conditions along the direction $x$ parallel to the walls. The position of the walls is controlled to ensure a constant normal stress $P^p$ and a constant mean shear velocity $\dot \gamma$. The particle and wall dynamics are integrated using a Verlet algorithm. These discrete elements are coupled to a density matched fluid, described as a slowly varying continuum phase. The hydrodynamical fluctuations of the pores \cite{Chareyre12} are neglected. The fluid velocity $\boldsymbol{u}^f(y)$ and the fluid shear stress $\boldsymbol{\sigma}^f(y)$ profiles are determined by averaging the equations governing the motion of the fluid over $x$ and over time $t$, as proposed in \cite{DAC12} (see Supplemental Material). 

The particles are submitted to four types of forces. (i) Upon contact, they interact with a viscoelastic force and with a Coulomb friction for relative tangential motion between particles at contact \cite{Cundall79,daCruz05,Luding06}. The model used for particle-particle interactions is identical to that proposed by Luding \cite{Luding06}. Quantities used in the model are expressed in terms of the grain density $\rho$, of the applied pressure $P^p$, and of the mean grain diameter $d$. In this system of units, the normal spring constant is chosen sufficiently large (between $10^3$ and $10^4$) to reach the rigid regime in which the results do not depend on it. The Coulomb friction coefficient is chosen equal to $\mu_p=0.4$, except for the inset of Fig.~\ref{fig:mu_phi} which is obtained in the frictionless limit $\mu_p=0$. The other viscoelastic parameters are chosen to lead to a restitution coefficient small enough (between $0.1$ and $0.9$) to get results that do not depend on it (see  Supplementary Information). (ii) They are submitted to a viscous drag force  given by:
\begin{equation}
 \boldsymbol{f}^{\rm drag}_i= 3\pi \eta_f (\boldsymbol{u}^f(y_i) -\boldsymbol{u}^p_i),
\end{equation}
which involves the nonaffine particle velocity component, i.e. the fluid velocity $\boldsymbol{u}^f$ minus the particle velocity $\boldsymbol{u}^p$, and where $i$ is the particle label. This is based on the assumption that the particle based Reynolds numbers ${\rm Re}_p=\frac{\rho\,|\boldsymbol{u}^p-\boldsymbol{u}^f| d }{\eta_f}$ remains small. (iii) When the fluid presents a stress gradient, it exerts a resultant Archimedes force on the particle, which reads $\boldsymbol{f}_i^{\rm archi} =\phi (1-\phi)^{-1} \boldsymbol{f}_i^{\rm drag}$ (see Supplemental Material). (iv) Finally, when particles are separated by a lubrication film, we include the extra stress as an interparticle force mediated by the fluid \cite{C74}: 
\begin{eqnarray}
\boldsymbol{f}^{\rm lubr,n}_{ij}(h_{ij}) &=& - \frac{3}{{8}} \pi \eta_f d_{ij}  \frac{(\boldsymbol{u}_i-\boldsymbol{u}_j)\cdot \boldsymbol{n}_{ij}}{(h_{ij}+\delta)},\\
\boldsymbol{f}^{\rm lubr,t}_{ij}(h_{ij}) &=& -\frac{1}{{2}} \pi \eta_f  \ln\Big( \frac{d_{ij}}{2(h_{ij}+\delta)}\Big)  (\boldsymbol{u}_i-\boldsymbol{u}_j)\cdot \boldsymbol{t}_{ij} \nonumber, \\ 
\end{eqnarray}
where $h_{ij}$ is the gap between the particles labelled $i$ and $j$,  $d_{ij}=\frac{2d_i d_j}{d_i +d_j}$ is the effective grain diameter, $\boldsymbol{n}_{ij}$ and $\boldsymbol{t}_{ij}$ are the normal and tangential unit vectors between the grains.  $\delta$ is a regularization length, chosen equal to 5\% of particle diameter. In real suspensions, it can be either related to the slip length, to the grain roughness or to the scale over which grains are elastically deformed~\cite{Rognon11}. This lubrication interaction is truncated for $h_{ij}>(d_i+d_j)/4$. 
\begin{figure*}[ht!]
\begin{center}
\includegraphics[scale=1]{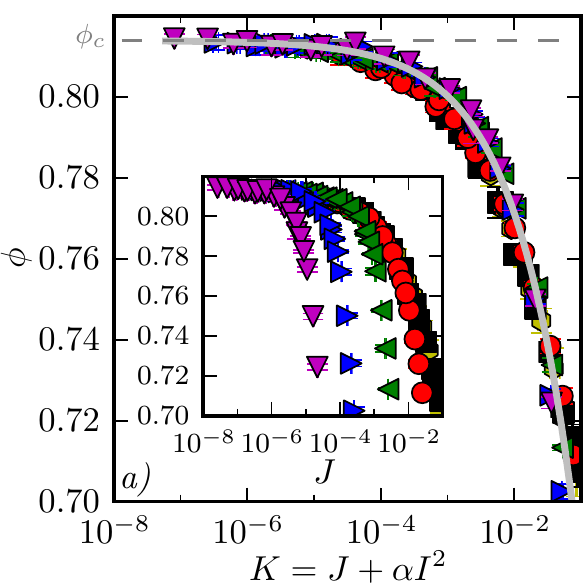}\includegraphics[scale=1]{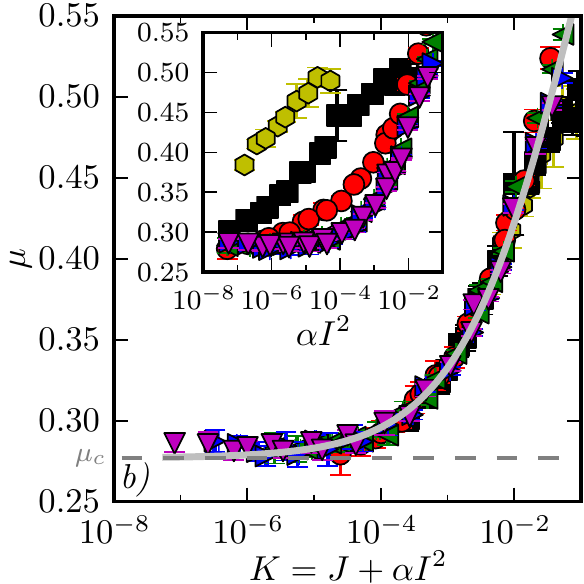}\includegraphics[scale=1]{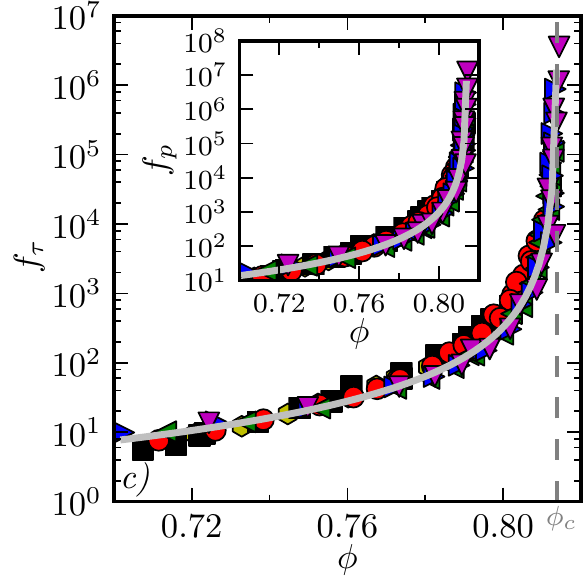}
\vspace{0 mm}
\caption{Simulated data. \textit{a)}: $\phi$ as a function of $K$ (inset as a function of $J$). \textit{b)}:  $\mu$ as a function of $K$ (inset as a function of $I^2$).  \textit{c)}: $f_\tau$ as a function of $\phi$ (inset shows $f_p$). All three Figures are for various $I/J$ with the same color coding as in Figure \ref{fig:mu_phi}. Solid lines corresponds to fits according 
to Eqs. 6,7, $f_\tau = \mu/K $ and $f_p = 1/K$. Dashed lines show the critical values.}
\label{fig:phi_K_mu_K}
\vspace{0 mm}
\end{center}
\end{figure*}

As the fluid is described as a continuum phase in a steady state, inertial effects and nonaffine effects are entirely ascribed to the particle phase. This means that the density $\rho$ only appears in the equation of motion for the grains and includes the added-mass effect.

As obtained for dense granular flows \cite{daCruz05}, the simulation is insensitive to microscopic parameters  provided that the grains are hard enough. The state of the system is then characterised by the two dimensionless numbers $I$ and $J$. In the following, we will rather use the Stokes number $I^2/J=\rho\dot \gamma d^2 /\eta_f$ and the rescaled confining pressure $I/J=\sqrt{\rho P^p}d/\eta_f$.

{\it Transition from viscous to inertial regime.~--} Fig.~\ref{fig:DispM} presents simulation results obtained at the same volume fraction $\phiÊ\simeq 0.78$ by varying the rescaled confining pressure $I/J$, where $I$ is typically varied from $10^{-3.5}$ to $10^{-0.5}$. It compares the contributions to the dissipated power of the different forces acting on the bulk of the suspension. This dissipation is balanced by the energy brought through the boundaries of the element of suspension considered. While the dissipation due to the drag force is dominant at small $I/J$, the dissipation in the contacts becomes dominant at large $I/J$ and the system resembles a dry granular flow (within the inclusion of the added mass effect inside the density $\rho$). The system therefore presents a transition from a viscous to an inertial regime, controlled by the rescaled pressure. It can be seen that the dissipation in the fluid, both in the pores and in the lubrication films, gives a subdominant contribution and varies like the contribution due to the drag force. In the following, we will therefore focus on results obtained without the lubrication forces. 

Looking at the inset of Fig.~\ref{fig:DispM}, one observes that the friction coefficient $\mu$ defined as the ratio of the particle shear stress $\tau^p$ and confining pressure $P^p$ remains constant across the transition. This means that, at fixed $\phi$, the shear stress is controlled by pressure, with a multiplicative factor insensitive to the nature of the dissipation mechanisms. Fig.~\ref{fig:mu_phi} shows the friction coefficient $\mu$ of the system as a function of the volume fraction $\phi$ for different values of the number $I/J$. A good data collapse is obtained, when $I/J$ is changed over five decades, showing that $\mu$ is a sole function of $\phi$. Moreover, the data obtained for frictional ($\mu_p=0.4$) and frictionless ($\mu_p=0$) particles fall on the same master curve \cite{daCruz05} and differ only by the values of $\phi_c$ and $\mu_c$.

{\it A single rheology across the transition.~--} It has been recently argued that, in the viscous quasistatic limit, trajectories are mostly controlled by geometric effects close to the jamming point, and do not depend much on the nature of the mechanisms dissipating energy \cite{Andreotti12}. We hypotheses here that the paths along which particles move do not vary much across the viscous/inertial crossover. In the viscous quasistatic limit, it was shown that {\it nonaffine} displacements control the enhanced dissipation close to jamming and take place over a time-scale vanishing as $\dot \gamma^{-1}\,(\phi_c-\phi)^{\beta/2}$, where $\beta\simeq 2$ is the divergence exponent of the stresses \cite{Andreotti12,Lerner12}. In the inertial regime, microscopic rearrangements take place over an inertial time scale $d \sqrt{\rho/P^p}$ \cite{daCruz05,GDRMidi}, over which we assume energy is dissipated. Assuming further that these two time scales are proportional to each other, we find that the stresses must also diverge as $(\phi_c-\phi)^{-\beta}$ in this regime.

Under the assumption that, for a given volume fraction $\phi$, the dissipation due to viscous effects and that due to grain binary interactions can simply be added, particle shear stress and confining pressure can be written as sums of linear (viscous) and quadratic (Bagnold) terms in $\dot{\gamma}$ \cite{GDRMidi,MS09,BGP11,Cassar05}:
 \begin{eqnarray}
\tau^p = f_\tau(\phi) \left(\eta_f \dot{\gamma} +\alpha \rho d^2 \dot{\gamma}^2\right), \label{taup} \\
P^p =  f_p(\phi) \left(\eta_f \dot{\gamma} +\alpha \rho d^2 \dot{\gamma}^2\right). \label{pp}
\end{eqnarray}
The best fit of $\mu$ and $\phi$, functions of $I$ and $J$, give a constant value of $\alpha = 0.635 \pm 0.009$ [Figs.~\ref{fig:phi_K_mu_K}(a),(b)]. Indeed, we obtain a collapse of all data when  $\mu$ and $\phi$ are plotted against $K=J+\alpha I^2$. This supports our above hypothesises. Consistently, expressions (\ref{taup}) and (\ref{pp}) give the two relations $\phi = f_p^{-1}(1/K)$ and $\mu = f_\tau(\phi)/f_p(\phi)$. Following empirical expressions proposed for $\phi$ and $\mu$ as functions of $I$ or $J$ in the cases of dry granular flows and dense suspension respectively \cite{daCruz05,GDRMidi,BGP11}, we can generalize them using the number $K$ as
\begin{eqnarray}
\label{eq:phik}
\phi(K)=\phi_c-b \sqrt{K}, \label{PhiofK} \\
\mu(K) = \mu_c + \frac{\mu_F-\mu_c}{1+\sqrt{K_0/K}}. \label{MuofK}
\label{eq:muk}
\end{eqnarray}
where $\phi_c = 0.8139 \pm 0.0003$ is the jamming volume fraction. The constants $b$, $\mu_c$, $\mu_F$, and $K_0$ are specific to the considered system. Here we find, $b=0.412\pm0.006$, $\mu_c=0.277\pm0.001$, $\mu_F=0.85\pm 0.01$ and $\sqrt{K_0}=0.29\pm0.01$.
Combining the two constitutive laws we finally get
\begin{equation}
\mu(\phi) = \mu_c + \frac{\mu_F-\mu_c}{1+\sqrt{K_0 b^2 / (\phi_c-\phi)^2}}.
\label{eq:muphi}
\end{equation}
This expression is in good agreement with the data displayed in Fig.~\ref{fig:mu_phi}. Furthermore, as $f_\tau = \mu/K $ and $f_p = 1/K$, these two functions are predicted to diverge close to the jamming point as $(\phi_c-\phi)^{-2}$, as a consequence of Eq.~(\ref{PhiofK}). This behavior is also very well supported by our data, as seen in Fig.~\ref{fig:phi_K_mu_K}(c).

We have run simulations in which lubrication interactions between the grains are taken into account. They  do not affect the qualitative results described above but slightly change the values of the constants. In particular, the exponent of the diverging behavior of both functions $f_\tau$ and $f_p$ is unchanged. This contradicts the claim of \cite{MS09} that the divergence would be in $(\phi_c-\phi)^{-1}$ when lubrication forces are present.

{\it Discussion.~--} The above analysis shows a crossover from viscous to inertial flow at a Stokes number $I^2/J = \dot{\gamma}d^2 \rho_s/\eta_f \simeq 1/\alpha$. The suspension is therefore found to present shear thinning at controlled granular pressure or shear thickening at controlled volume fraction, when $I^2/J$ goes beyond this value. In the experiments of Boyer \textit{et al.}, the maximum value of the Stokes number can be estimated as $10^{-3}$. This value is far below the inertial regime, and, consistently, all their rheological data collapse when using $J$ as the single dimensionless parameter \cite{BGP11}. By contrast, Fall \textit{et al.} report in their experiments a crossover between the two regimes, at a Stokes number of $2\,10^{-3}$ \cite{Fall2010}. This value is three to four orders of magnitude lower than the predictions of our simulations. We hypothesize that this effect may result from nonlocal effects, as the base flow is heterogenous. The dominant influence of nonlocality has previously been observed in other heterogenous flows of dense suspensions \cite{BLLC10}, emulsions \cite{Goyon2008}, and granular systems \cite{Andreotti2007,Reddy11,Pouliquen09}. Our setup is insensitive to nonlocal effects as all studied quantities are homogenous in the central part of our shear cell. Nevertheless, many flows are heterogenous and it would be important to understand nonlocality in order to rationalize even these. 

Newtonian fluids exhibit a transition from laminar to turbulent flow controlled by the Reynolds number based on the size of the flow and on the suspension viscosity. It is unlikely that dilute or even moderately concentrated suspensions would be an exception to this rule. As the jamming transition is approached ($\phi\to\phi_c$), the suspension viscosity diverges so that the Reynolds number vanishes. The transition from the viscous to the inertial regime in dense suspension is thereby of a different nature than the transition from laminar to turbulent flow.
In the former, both the Newtonian and Bagnoldian regimes are controlled by particle fluctuations with respect to the affine field. These fluctuations are controlled by the Stokes number, which is based on the grain diameter and the fluid viscosity rather than the suspension viscosity. 
Further studies are needed to investigate the transition from the inertial regime to the turbulent regime when the particle volume fraction is lowered.

In this Letter, we have shown that the Newtonian rheology of suspensions can be unified with the Bagnoldian shear-thickening regime for vanishing temperature. As pointed out recently by Ikeda \textit{et al.} \cite{Ikeda12}, thermal and athermal suspensions seem physically distinct, making a unified description of glass and jamming transitions unlikely. Future studies will have to explain the difference in nature (if any) between mechanically induced fluctuations (i.e. nonaffine motion) at zero temperature and thermal fluctuations.

We thank Orencio Dur\'an for support with the DEM code. We thank E. Cl\'ement, Y. Forterre, J. Kurchan, A. Lindner,O. Pouliquen and M. Wyart for discussions. This work is funded by ANR JamVibe.

\section{Supplemental Material)}
\subsection{Solving the two-phase hydrodynamics model}
In the presence of particles occupying a volume fraction $\phi$, the hydrodynamics is described by the two-phase flow Reynolds averaged Navier-Stokes equations
\cite{Jackson2000}. For the fluid phase we have:
\begin{equation}
\rho_f (1-\phi) \frac{D \bold{u}^f}{Dt} = \nabla \cdot \boldsymbol{\sigma}^f - \bold{F} + \rho_f (1-\phi) \bold{g}.
\end{equation}
Where $\rho_f$ denotes the density of the fluid phase, $\bold{g}$ the gravity field, $\boldsymbol{\sigma}^f$ the stress tensors
for fluid phase (with $\sigma^f_{ij} = -p^f\delta_{ij} + \tau^f_{ij}$ ), $\bold{F}$ the coupling term between the fluid and particles, and where $\frac{D}{Dt}$ is the material derivate given by:
\begin{equation}
\frac{D \bold{u}^f}{Dt}= \frac{\partial \bold{u}^f}{\partial t} +  (\bold{u}^f \cdot \nabla) \bold{u}^f.
\end{equation}
Our model assumes a newtonian fluid ($\tau^f_{xy} = \eta_f \partial u^f_x / \partial y$) in steady-state and without inertia. Assuming further that we are at zero gravity
reduces Eq.(1) to:
\begin{equation}
0 = \nabla \cdot \boldsymbol{\sigma}^f  - \bold{F}.
\label{Fcoup}
\end{equation}
The coupling term is given by the sum of the drag forces $\boldsymbol{f}^{\rm drag}_k$ and the Archimedes forces $\boldsymbol{f}^{\rm archi}_k$ over all particles labelled $k$ in a given volume $dV$:
\begin{equation}
\bold{F} = \frac{1}{dV} \sum_{k\in dV} \big( \boldsymbol{f}^{\rm drag}_k + \boldsymbol{f}^{\rm archi}_k \big).
\label{Fcoup2}
\end{equation}
The Archimedes force is given by $\boldsymbol{f}^{\rm archi}_k=V_p \nabla  \cdot \boldsymbol{\sigma}^f_k$, where $V_p$ is the particle volume and
$\sigma^f_k$ is the fluid stress exerted on grain $k$. 
Applying Eqs.~(\ref{Fcoup}) and (\ref{Fcoup2}) over a single grain with $dV\simeq V_p / \phi$ one obtains:
\begin{equation}
 \boldsymbol{f}^{\rm archi}_k \simeq \frac{\phi}{(1-\phi)} \boldsymbol{f}^{\rm drag}_k.
\end{equation}
Which then leads to
\begin{equation}
\bold{F} =  \frac{1}{(1-\phi)} \frac{1}{dV} \sum_{i\in dV}  \boldsymbol{f}^{\rm drag}_i = \frac{1}{(1-\phi)} \bold{F}^{\rm drag},
\end{equation}
in agreement with Jackson \cite{Jackson2000}. Our formalism applies the two-phase formalism on a single grain scale and thereby accounts for some hydrodynamical fluctuations
at this scale. \\
The fluid velocity field is found by sampling the $x$-component of the coupling term and integrating twice:
\begin{equation}
u_x^f(y) =  u_x^f(0)+\int^y_0 \frac{1}{\eta_f}  \Big(\int_0^h  \left<\bold{F}_x(y'')\right> dy'' + \tau^f_{0}\Big) dy'
 \end{equation}
Where $\left<\bold{F}_x(y)\right>$ is space- and time-averaged over a thin horizontal region and 200 time-steps. A new $u_x^f$ fluid profile is calculated every 200th time-step. To preserve no-slip boundary condition at the walls we added at each time a constant stress, $\tau^f_{0}$, such to ensure the no-slip boundary condition.
The fluid profile
was continuously iterated to convergence in the simulations using the two-phase coupling term as a feed-back mechanism for the fluid, with a damping mechanism which makes use of the average of the $10^4$ last fluid profiles rather than the instantaneous fluid velocity profile. The numbers in our averaging protocol are chosen in such a way to have a good balance between velocity update/convergence and stability. We checked that other averaging numbers give the same results.
\subsection{Viscoelastic parameters }
Units used in the model are expressed in terms of the grain density ($\rho$), granular pressure ($P^p$), and mean grain diameter ($d$). With these scales $k_n = 10^3,10^4 \, (P^p)$ (granular pressure), $k_t=0.5 \, k_n$ (tangential spring constant),  $\beta_n\approx1.33,4.20\, (\sqrt{P^p \rho d^2})$ (normal damping), and $\beta_t\approx0.94,2.97\,  (\sqrt{P^pÊ\rho d^2})$ (tangential damping) for $\mu_p=0.4$. This yields a coefficient of restitution $e\simeq 0.9$. For $\mu_p=0$ runs with $k_n = 10^3 \, (P^p)$ and $\beta_n\approx1.33, 23.43 \,  (\sqrt{P^pÊ\rho d^2})$ were also performed, corresponding to $e\simeq 0.9, 0.1$. Especially at high shear rates (i.e. high $I$) for the frictionless grains one finds a dependence on $\mu$ vs $I$ or $\phi$ on the value of the coefficient of restitution $e$. We choose to only show data points which are independent of the choice of $e$ (within the interval 0.1 to 0.9). This
roughly corresponds to points with two or more contacts in average per grain in the frictionless case. We see this as a signature of being in a liquid-like rather than in a gaseous regime.

\end{document}